\documentclass[runningheads]{llncs}
\usepackage[T1]{fontenc}
\usepackage{subfigure}
\usepackage{amsmath}
\usepackage{tikz}
\usetikzlibrary{shapes.geometric, arrows, positioning, fit}
\usepackage{enumitem}
\usepackage{array}
\usepackage{hyperref}
\usepackage{lscape} 
\usepackage{longtable} 
\usepackage{tikz}
\usetikzlibrary{positioning,shapes,shadows,arrows.meta,calc}
\usepackage{pgfplots}
\pgfplotsset{compat=1.18}
\usepackage{graphicx}
\usepackage{geometry}
\usepackage{colortbl}
\usepackage{xcolor}
\usepackage{svg}
\usepackage{float}

\begin{document}
\title{Analyzing Social Networks of Actors in Movies and TV Shows}
%
%
\author{Sarthak Giri\inst{1,*} \and
Sneha Chaudhary\inst{1} \and
Bikalpa Gautam\inst{2}}
\authorrunning{Giri et al.}
%
\institute{Software Engineer, ION Group, Noida, Uttar Pradesh, India \and
Research Officer, Singapore Institute of Manufacturing Technology, 2 Fusionopolis way, Singapore
*\email{girisarthak84@gmail.com}\\
}
\maketitle          
\begin{abstract}
The paper offers a comprehensive analysis of social networks among movie actors and directors in the film industry. Utilizing data from IMDb and Netflix, we leverage Python and NetworkX to uncover valuable insights into the movie industry's intricate web of collaborations. Key findings include identifying the top actors and directors in the OTT sector, tracking the rise of movies on OTT platforms, and analyzing centrality measures for actors. We also explore the hidden patterns within the movie data, unveiling the shortest paths between actors and predicting future collaborations. Cluster analysis categorizes movies based on various criteria, revealing the most insular and liberal clusters and identifying crossover actors bridging different segments of the industry. The study highlights that actors predominantly collaborate within language groups, transcending national boundaries. We investigate the degree of isolation of Bollywood from global cinema and identify actors working across world clusters. The project provides valuable insights into the evolving dynamics of the film industry and the impact of OTT platforms, benefiting industry professionals, scholars, and enthusiasts.

\end{abstract}
\renewcommand{\thefootnote}{\textit{ab}} 
\footnotetext{The following abbreviations are used in this manuscript:\\
 \begin{tabular}{@{}ll}
OTT & Over-The-Top\\
SNA & Social Network Analysis\\
IMDb & Internet Movie Database\\
EDA & Exploratory Data Analysis\\
NLP & Natural Language Processing\\
QoE & Quality of Experience\\
PCA & Principal Component Analysis\\
ANN & Artificial Neural Network\\
\end{tabular}}
\vspace{-1cm}
\section{Introduction}
The rapid rise of Over-the-top (OTT) streaming services alongside the long-established film industry is driving a major and transformative shift in the entertainment landscape \cite{b1,b2,b3,b4}. A lot of research has been done already that studied the performance of OTT platforms and their impact on the market \cite{d1,d2,d3}. For performers and artists, this is a critical time that offers them a unique chance to engage with audiences around the world \cite{d4,d5}. In this digitally infused era, a strong online presence has become essential to success, and the OTT sector has become a vital platform for actors and filmmakers to flourish on.
But in the fiercely competitive OTT space, the complex web of interactions and relationships between market players becomes a critical determinant of success \cite{b5,b6}. A thorough understanding of the interactions between actors, directors, producers, and the industry at large is imperative in this digital age. The implications of the entertainment industry's ongoing digital age adaptation are wide-ranging, encompassing aspects such as cross-industry dynamics, collaboration patterns, and the changing nature of actor relationships.

SNA, a data science discipline, provides a powerful framework for investigating these interactions by leveraging the power of networks and graph theory \cite{d6,d7,d8,d9}. SNA has a long history of revealing hidden patterns in a variety of fields, from determining how illnesses spread to tracking the dissemination of ideas. In order to investigate the intricate connections within the traditional film industry and the OTT industry, we are utilizing SNA for movie network analysis. Our data collection includes publicly available sources, with platforms like IMDb and Netflix functioning as crucial repositories. Using the NetworkX package and the Python programming language, the analysis allows us to carefully examine this data and reveal important insights.

Additionally, our study explores the centrality metrics of different players, illuminating their roles within the networks and exposing the dominant figures that influence the sector. We investigate the idea of the shortest path between players, revealing hidden relationships that specify the paths leading to cooperation. Furthermore, in order to predict future partnerships, we utilize advanced link prediction approaches like resource allocation, Jaccard cosine similarity, common neighbour analysis, and Adamic Adar index. The study explores the fascinating world of clusters within the network, exposing subclusters within clusters, raising important questions about Bollywood's level of exclusion from the international film industry, and identifying performers who work across different countries.
This research project is motivated by the investigation of these interrelated networks and the complex relationships that characterize the cinema and over-the-top (OTT) industries. The following sections of the paper include a literature survey, methodology, results, and conclusion, as well as future work.

\section{Related Works}

The related literature provides valuable insights into sentiment and social network analysis in movie data. However, our research builds on these studies by expanding the scope, applying deeper network analysis, and using more extensive datasets, as summarized in Table \ref{tab:related_work}. 

\section{Keyword Network Analysis of Actor Collaborations in Movies and OTT Platforms} 

We conducted a keyword co-occurrence analysis to examine the evolving dynamics of social networks within the movie and television industries. Our dataset comprised research papers sourced from the Web of Science, specifically filtered for relevance to ``Social Network Analysis in Movies.'' This allowed us to identify key contributors and emerging trends in the field.

\begin{figure}[h]
  \begin{center}
    \includegraphics[width=1\textwidth, height=1.25\textheight, keepaspectratio]{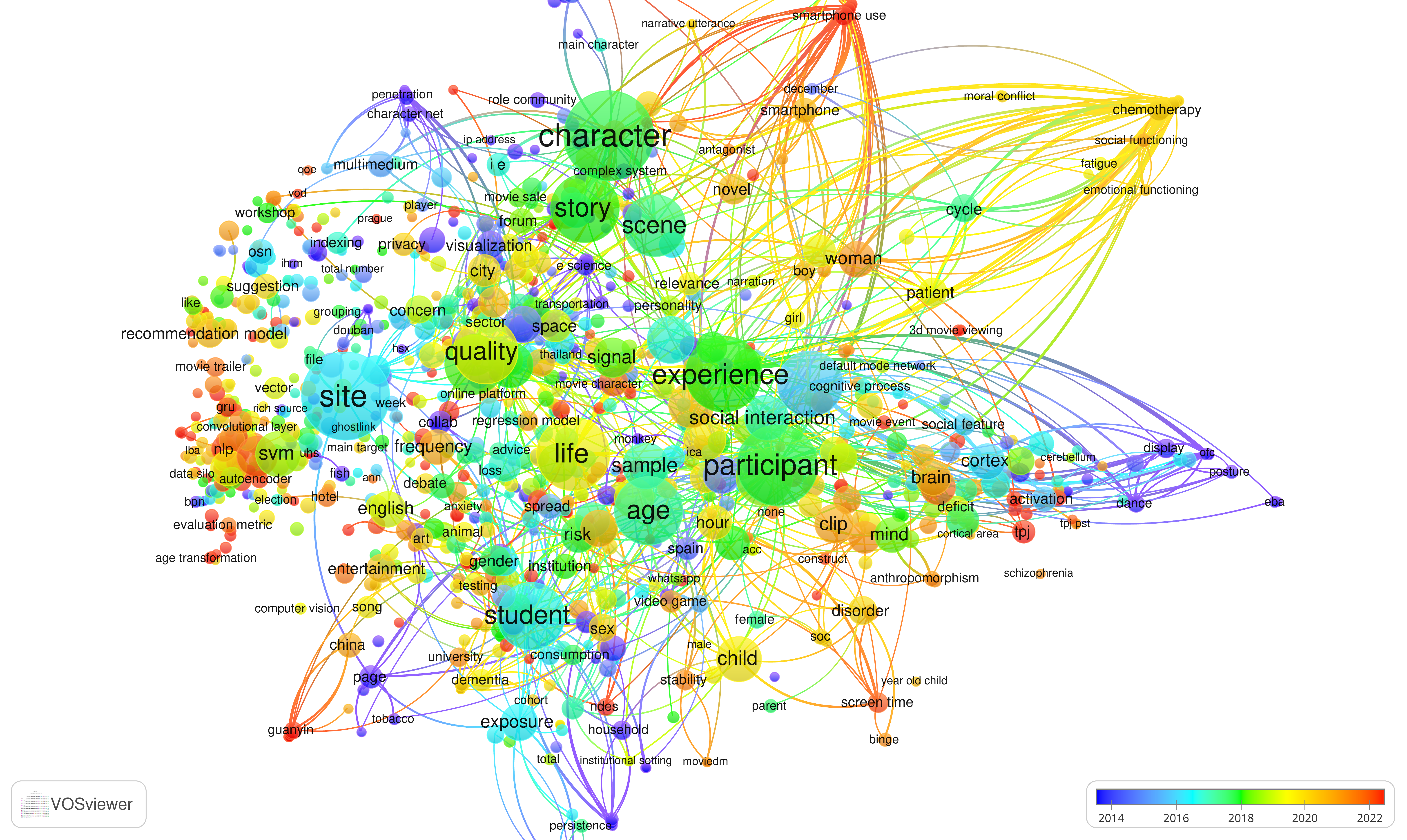}
    \caption{Keyword co-occurrence network illustrating key themes in social network analysis of actors and directors, and the role of digital platforms and character centrality in shaping film and OTT collaborations between 2014-2022}
    \label{fig:1}
  \end{center}
\end{figure}

The resulting keyword co-occurrence network, visualized using VOSviewer (see Fig. \ref{fig:1}), provides insights into dominant research areas such as ``character,'' ``story,'' and ``experience.'' These terms reflect the role of actors and directors within the social fabric of the industry. Notably, the emergence of keywords like ``smartphone use'' and ``OTT streaming services'' indicates a shift towards digital platforms, such as Netflix and Amazon Prime, which are reshaping traditional cinematic practices. These platforms foster new forms of audience interaction and story development, influencing real-time engagement and narrative progression. The prominence of terms like ``quality'' and ``experience'' highlights the intersection of technical film critique and audience reception. These social dynamics, particularly through OTT services, directly influence viewers' engagement with content and, consequently, the success of films. The color-coded temporal analysis, spanning from 2014 to 2022, reveals a growing focus on ``multimedium'' and ``smartphone use,'' signaling key directions for future research. This keyword co-occurrence network not only reflects current research priorities but also provides a framework for investigating the ongoing transformation within the film industry. These emerging areas suggest that the role of digital platforms in shaping social networks within the industry warrants further exploration. It underscores the shifting paradigm of the film industry, where traditional collaborations meet digital disruption.

\begin{landscape}
\begin{longtable}[htbp]{|p{3.5cm}|p{8cm}|p{10cm}|}
\hline
\textbf{Author(s)} & \textbf{Contributions} & \textbf{Limitations/How Our Research Addresses Them} \\ \hline

Dangi et al. \cite{c1} & Framework for sentiment and pattern analysis in movie data. & Lacked community detection, centrality, and database expansion beyond movies. We include OTT content and broader network metrics. \\ \hline

Hodeghatta \cite{c2} & Twitter sentiment analysis of Hollywood movies. & Focused on Hollywood only. Our research covers a wider database, including OTT platforms. \\ \hline

Weng et al. \cite{c3} & Semantic movie analysis using role-based social networks. & Applied to only three types of storylines. We extend the analysis to more complex movie narratives and databases. \\ \hline

Ha et al. \cite{c4} & CosMovis for sentiment relationships in movie reviews. & Inefficient for large networks. Our system handles large datasets more effectively. \\ \hline

Weng and Chu \cite{c5} & RoleNet: character exploration in stories. & Limited databases affect accuracy. We use larger, more diverse datasets for improved results. \\ \hline

Rathor and Prakash \cite{c6} & Sentiment analysis using IMDB reviews. & Limited to IMDB. Our research includes both movie and OTT databases, covering actor-movie/series networks. \\ \hline

Li and Lin \cite{c7} & Key player identification in social networks. & Focused on egocentric networks. Our work applies these methods to actor networks with broader metrics. \\ \hline

Lee et al. \cite{c8} & ActRec: Actor recommendation using word embeddings. & Focused on actor-role recommendation. Our research delves into actor networks using advanced analysis. \\ \hline

Hu et al. \cite{c9} & Movie and actor recommendations based on YAGO and IMDB. & Limited to movie recommendations. Our analysis applies to actor networks, using comprehensive datasets. \\ \hline

Liu et al. \cite{c10} & SMAS: Social media analysis for box office prediction. & Focused on revenue prediction. Our study extends to actor network analysis across movies and OTT platforms. \\ \hline

Landherr et al. \cite{c11} & Reviewed centrality measures in social networks. & Centrality measures are not applied to actor networks. Our research implements these metrics for actor network analysis. \\ \hline

Wang et al. \cite{c12} & Heterogeneous graph attention network with sentiment markers. & Focused on graph representation. Our research performs extensive network analysis beyond representation. \\ \hline

Yuliana et al. \cite{c13} & SNA applied to an online fan group. & Limited to online fan analysis. Our research covers larger real-world datasets in movies and OTT content. \\ \hline

Lewis et al. \cite{c14} & Social networks and cultural tastes on Facebook. & Focused on changing traits like hobbies. We use a constant movie dataset for more stable insights. \\ \hline

Tang et al. \cite{c15} & TAP model for social influences in large networks. & Focused on connections and relationships. Our research extends to centrality measures and community detection. \\ \hline

\caption{Comparison of related works with our research.}
\label{tab:related_work}
\end{longtable}
\end{landscape}

\vspace{-1cm}
\section{Proposed Architecture} 

\begin{figure}
  \centering
  \includegraphics[width=\textwidth]{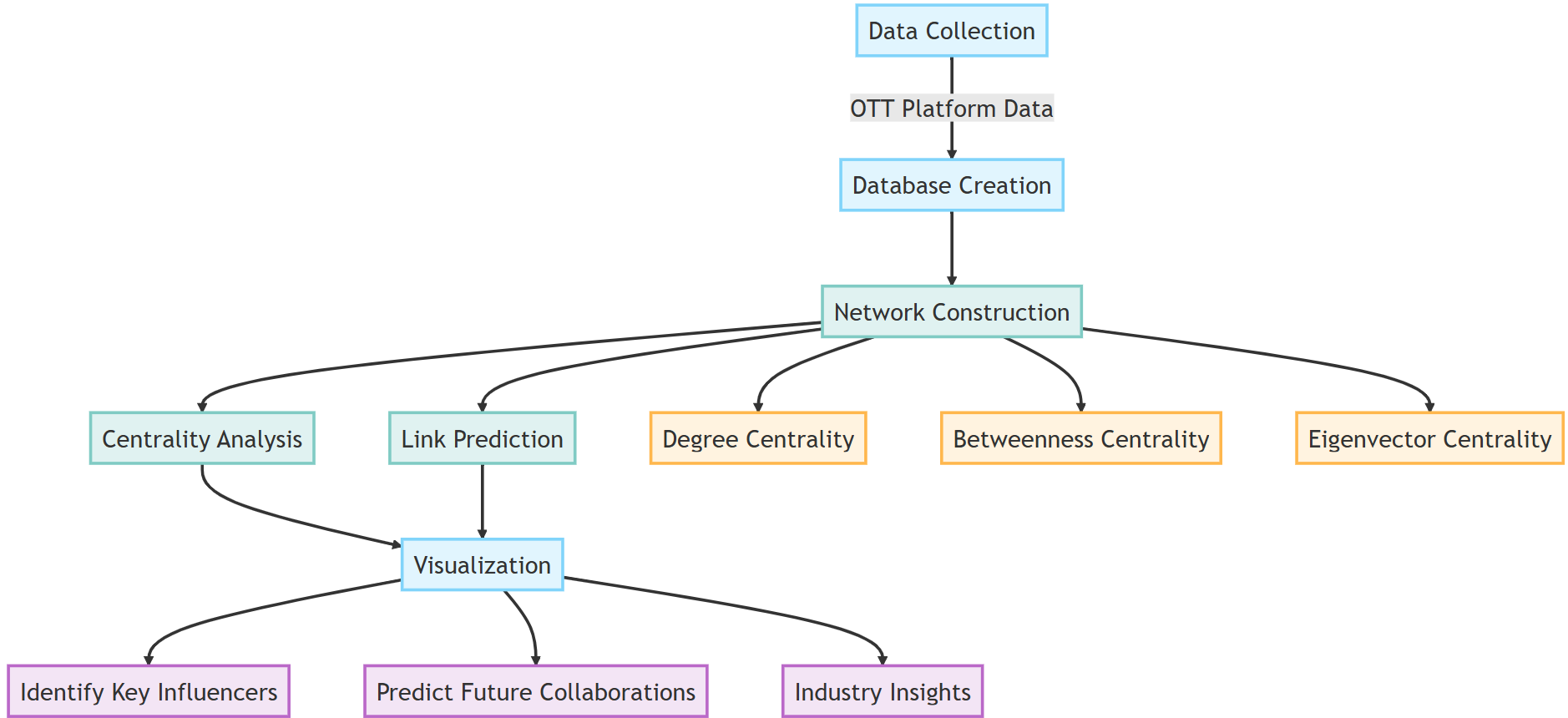} 
  \caption{Proposed architecture outlining the workflow for social network analysis within OTT platforms, incorporating data collection, network construction, and centrality analysis to identify key influencers.}
  \label{movieshows}
\end{figure}

This proposed architecture given in Fig. \ref{movieshows} outlines a comprehensive approach to conducting social network analysis within the OTT and movie industries. It encompasses data collection from diverse sources, like Kaggle\textsuperscript{\ref{fnk
}} and IMDb\textsuperscript{\ref{fni
}}. By leveraging the NetworkX library in Python, the architecture facilitates network construction to visualize relationships between various actors. Centrality measures, such as degree, betweenness, and eigenvector centrality, are applied to identify influential actors who play pivotal roles in connecting disparate groups and influencing information flow. Additionally, link prediction algorithms like the Jaccard coefficient calculate the likelihood of potential collaborations, with the caveat that some probabilities may exceed 1 due to high connectivity within the network. Visualization tools further enhance the understanding of actor relationships, while community detection methods, such as the Louvain method, reveal clusters of collaboration and shared interests among actors. Ultimately, this analysis aims to provide valuable insights into viewer preferences and emerging trends, identifying potential future leaders and collaborations in the movie industry landscape.

\renewcommand{\thefootnote}{\textit{c}} \footnotetext[1]{\label{fnk
}\url{https://www.kaggle.com/datasets/shivamb/netflix-shows}} \renewcommand{\thefootnote}{\textit{d}} \footnotetext[2]{\label{fni
}\url{https://developer.imdb.com/non-commercial-datasets/}}
 
\section{Methodology}

This paper's methodology involves the following sequential steps: Firstly, we will carefully choose a suitable database to establish the network. Subsequently, we will conduct preprocessing and culling of the database to eliminate undesirable data and entries that are not useful for our analysis. Next, we will construct a social network from this refined database, with a preference for representing it as an adjacency matrix. This matrix will serve as a foundational tool for our analysis. We will then engage in a series of operations on the matrix to draw meaningful conclusions, such as determining centralities and exploring various factors like identifying actors with the highest number of film credits, pinpointing actors with extensive co-star collaborations, and uncovering co-stars who have participated in the most acting partnerships.

\subsection{Dataset}\label{AA}

The study conducts a social network analysis of movie actors using data from Netflix, a leading global OTT platform, and IMDb, an open-source database rich in information on movies, series, and entertainment figures. The dataset includes actor names, titles, release dates, and co-appearance details, which are crucial for building the social network. The OTT data is sourced from Kaggle and contains over 8,807 unique streaming titles, while the movie data is taken from IMDb.

\subsection{Modules}

In this research, we developed multiple interconnected modules to systematically analyze the collaboration patterns in the film and OTT industries. First, the Data Collection and Preprocessing module handled the input data, performing essential pre-processing and cleaning steps to create a structured dataset and a network graph for further analysis in Python. Then we had the Exploratory Data Analysis (EDA) module, which provided a basic overview of the dataset by revealing key patterns and trends within the OTT and movie industries. Through graphical visualizations, this module showcased critical metrics such as the number of movies released per year, average cast size, content type distribution, rating categories, and viewer preferences.

Following this, the Actor Collaboration Network module used the network graph to evaluate metrics such as the shortest path between any two actors and the likelihood of their collaboration based on the path length. By analyzing these connections, we gained insights into actor partnerships and identified actors with significant collaboration histories.

To further analyze influence within the network, the Influence and Centrality Analysis module applied several centrality measures. Degree Centrality measured the number of connections \cite{c16} each actor had, quantifying their collaborative activity, as captured by the equation:
\[
C_D(n_i) = \frac{x_{i+}}{g-1}
\]
This metric highlighted actors with a high number of collaborations. Betweenness Centrality \cite{c16}, represented by the equation:
\[
C_B(n_i) = \frac{C_B(n_i)}{(g-1)(g-2)}
\]
evaluated actors who bridged different clusters, such as actors working across Hollywood and British cinema. Closeness Centrality \cite{c16}, given by:
\[
C_C(n_i) = \frac{g-1}{\sum_{j=1}^{g} d(n_i, n_j)}
\]
measured how close an actor was to others in the network, providing a metric for accessibility and potential collaboration opportunities. Eigen Centrality \cite{c16}, described by:
\[
x_v = \frac{1}{\lambda} \sum_{t \in M(v)} x_t = \frac{1}{\lambda} \sum_{t \in G} a_{v,t} x_t
\]
went beyond simple connections, evaluating actors based on their ties to other well-connected figures, highlighting influential actors and directors within the network.

In order to anticipate future collaborations, the Collaboration Prediction module applied algorithms such as Common Neighbour \cite{c17}, Jaccard Coefficient \cite{c18,c19}, Resource Allocation \cite{c20}, Adamic-Adar Index \cite{c21}, and Preferential Attachment \cite{c22}. For example, the Jaccard Coefficient measured the similarity of actors’ collaborative networks through the formula:
\[
\sigma_{\text{Jaccard}}(v_i, v_j) = \frac{|N(v_i) \cap N(v_j)|}{|N(v_i) \cup N(v_j)|} 
\]
These algorithms enabled the prediction of potential future collaborations by examining the shared connections between actors and directors, offering a probabilistic outlook on future industry partnerships.

Next we had Community Detection Module, where we employed Louvain Clustering to group actors into distinct communities based on their collaborative patterns. This clustering technique identified clusters of actors who frequently work together, forming subcommunities within the broader industry network. By analyzing these subclusters, we discovered smaller, tightly knit groups of actors with recurring collaborations. Furthermore, by extending this analysis to examine cross-cluster connections, we identified key actors who serve as bridges between different sections of the industry. These bridging actors facilitate the flow of talent and ideas across both regional and international film markets, highlighting influential individuals who contribute to cross-market collaborations. It helped us analyze how the clusters evolved overtime.

\begin{center}
    \begin{tikzpicture}[
    actor/.style={
        line width=0.8pt,
        minimum size=0.8cm
    },
    usecase/.style={
        draw=gray!50,
        line width=0.4pt,
        ellipse,
        fill=white,
        drop shadow={shadow xshift=0.5pt, shadow yshift=-0.5pt},
        minimum width=1.8cm,
        minimum height=0.5cm,
        align=center,
        font=\sffamily\scriptsize,
        inner sep=1pt
    },
    maincase/.style={
        draw=gray!50,
        line width=0.6pt,
        ellipse,
        fill=blue!15,
        drop shadow={shadow xshift=0.8pt, shadow yshift=-0.8pt},
        minimum width=2cm,
        minimum height=0.6cm,
        align=center,
        font=\sffamily\footnotesize\bfseries,
        inner sep=2pt
    },
    extends/.style={
        -Stealth,
        dashed,
        gray!80,
        thin,
        shorten >=1pt,
        shorten <=1pt
    }
]
\begin{scope}[shift={(-5.5,0)}]
    \draw[line width=0.8pt] (0,0) circle (0.15);
    \draw[line width=0.8pt] (0,-0.15) -- (0,-0.6);
    \draw[line width=0.8pt] (-0.3,-0.3) -- (0.3,-0.3);
    \draw[line width=0.8pt] (0,-0.6) -- (-0.2,-0.9);
    \draw[line width=0.8pt] (0,-0.6) -- (0.2,-0.9);
\end{scope}

\node[maincase] (prelim) at (-2,2.5) {Preliminary\\Analysis};
\node[maincase] (temporal) at (-2,1.0) {Analysing\\Connections};
\node[maincase] (demographic) at (-2,-0.5) {Connectivity and\\Centrality Analysis};
\node[maincase] (hotspot) at (-2,-2.0) {Link Prediction};
\node[maincase] (community) at (-2,-3.5) {Community\\Detection};

\begin{scope}[xshift=3cm]
    \node[usecase] (hotspot1) at (0,3.3) {Movies per year};
    \node[usecase] (hotspot2) at (0,2.7) {Movies per cast size};
    \node[usecase] (hotspot3) at (0,2.1) {Movies per actor};
    
    \node[usecase] (shortpath) at (0,1.36) {Shortest path among\\two actors};
    \node[usecase] (seasonal) at (0,0.64) {Acting partnerships};
    
    \node[usecase] (density1) at (0,0) {Degree centrality};
    \node[usecase] (density2) at (0,-0.6) {Closeness centrality};
    
    \node[usecase] (common) at (0,-1.2) {Common neighbour};
    \node[usecase] (jaccard) at (0,-1.8) {Jaccard coefficient};
    \node[usecase] (resource) at (0,-2.4) {Resource Allocation};
    \node[usecase] (pref) at (0,-3.0) {Preferential Attachment};
    \node[usecase] (adamic) at (0,-3.6) {Adamic-Adar Index};
    
    \node[usecase] (louvain) at (0,-4.2) {Louvien Clustering};
    \node[usecase] (commevo) at (0,-4.8) {Community evolution};
\end{scope}

\draw[extends, line width=0.8pt] (-5.5,-0.3) -- (prelim);
\foreach \dest in {temporal,demographic,hotspot,community}
    \draw[extends] (-5.5,-0.3) -- (\dest);

\foreach \dest in {hotspot1,hotspot2,hotspot3}
    \draw[extends] (prelim) -- (\dest);
\foreach \dest in {shortpath,seasonal}
    \draw[extends] (temporal) -- (\dest);
\foreach \dest in {density1,density2}
    \draw[extends] (demographic) -- (\dest);
\foreach \dest in {common,jaccard,resource,pref,adamic}
    \draw[extends] (hotspot) -- (\dest);
\foreach \dest in {louvain,commevo}
    \draw[extends] (community) -- (\dest);

\end{tikzpicture}
\end{center}

\section{Results and Analysis}

\begin{figure}[H]
  \centering
  \begin{minipage}[b]{0.49\textwidth}
    \centering
    \includegraphics[width=\textwidth]{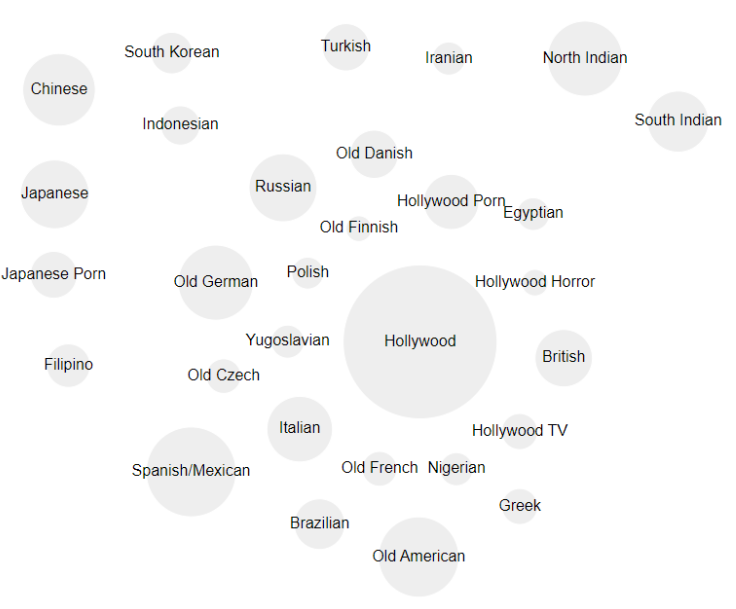}
    \caption{Initial raw clusters before applying interconnection frequency filters}
    \label{otttrends}
  \end{minipage}
  \hfill
  \begin{minipage}[b]{0.49\textwidth}
    \centering
    \includegraphics[width=\textwidth]{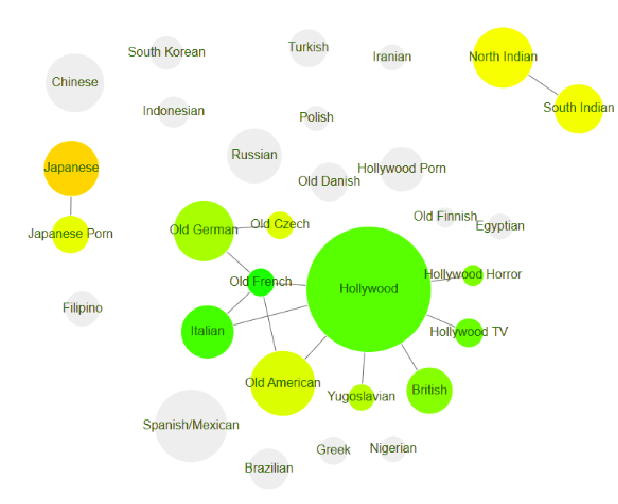}
    \caption{Visualization of cluster interconnection at 2\% frequency rate}
    \label{2percent}
  \end{minipage}
\end{figure}

Our research on actor collaboration networks in the traditional film industry and OTT platforms introduces several insights that challenge existing conclusions in the field. While Rathor and Prakash \cite{c6} examined movie sentiment analysis from user reviews, our network-based approach reveals the structural dynamics driving those sentiments. Using network analysis and centrality measures, we focused on top actors, including Anupam Kher, Takahiro Sakurai, and others, to identify their influence in bridging different groups. As shown by Table \ref{degreecentrality} and Table \ref{closenesscentrality}, degree centrality identified Kher and Sakurai as the most connected actors, highlighting their role in fostering cross-industry collaborations, while Fred Tatasciore and Maya Rudolph showed the highest closeness centrality, underscoring their importance as connectors between diverse clusters. Further, the analysis of the shortest path between actors exposed fascinating details of acting relationships, such as Robin Williams' connection to Angelina Jolie through movies like ``Dead Poets Society'' and ``Taking Lives''.
\vspace{-0.5cm}
\begin{table}[h]
  \centering
  \begin{minipage}{0.49\textwidth}
    \centering
    \small
    \renewcommand{\arraystretch}{1.5}
    \begin{tabular}{|p{1.5cm}|p{1.5cm}|p{1.5cm}|p{1.5cm}|}
      \hline
      Anupam Kher & Takahiro Sakurai & Yuichi Nakamura & Fred Tatasciore \\
      \hline
      0.00750 & 0.00677 & 0.006192 & 0.006134 \\
      \hline
    \end{tabular}
    \vspace{6pt} 
    \caption{Actors with highest betweenness centrality}
    \label{degreecentrality}
  \end{minipage}
  \hfill
  \begin{minipage}{0.49\textwidth}
    \centering
    \small
    \renewcommand{\arraystretch}{1.5}
    \begin{tabular}{|p{1.5cm}|p{1.5cm}|p{1.5cm}|p{1.5cm}|}
      \hline
      Fred Tatasciore & Fred Armisen & Anupam Kher & Yuichi Nakamura \\
      \hline
      0.255 & 0.2518 & 0.2415 & 0.2080 \\
      \hline
    \end{tabular}
    \vspace{6pt} 
    \caption{Actors with highest closeness centrality}
    \label{closenesscentrality}
  \end{minipage}
\end{table}\\ \vspace{-1.1cm}
\\
Our analysis further explored the formation and evolution of clusters, representing different movie industries, such as Hollywood, British, Spanish, Brazilian, and more. We observed how these clusters gradually evolved, with the
degree of constraints influencing the degree of intercluster interactions. At a 5\% interaction frequency, we witnessed clusters like Hollywood pairing with Hollywood Horror, TV, and British, indicating their propensity to collaborate only within their niche. As constraints were reduced, Hollywood expanded its reach, forging connections with British, German, French, Czech, Yugoslavian, and Italian actors, forming a substantial connected cluster. This evolution showcased the adaptability of the industry in transcending traditional boundaries and affiliations. Intriguingly, we found that actors primarily collaborate based on language rather than country, reflecting the significance of linguistic bonds within the industry.
\vspace{1cm}
\begin{table*} 
  \centering
  \small
  \renewcommand{\arraystretch}{1.5} 
  \setlength{\tabcolsep}{8pt} 
  \begin{tabular}{|p{2.5cm}|p{6.5cm}|}
    \hline
    \textbf{Probability of Connections} & \textbf{Actors} \\
    \hline
    0.84 & (Joseph Gordon, Gary Oldman), (Gary Oldman, Marion Cotillard) \\
    \hline
    0.63 & (Liam Neeson, Gary Oldman), (Tom Hardy, Tom Wilkinson), (Marion Cotillard, Rutger)\\
    \hline
    0.54 & (Liam Neeson, Aaron Eckhart), (Maggle G, Rutger Hauser) \\
    \hline
  \end{tabular}
  \vspace{6pt} 
  \caption{\label{linkprediction} Link prediction probabilities between actors in the OTT and movie industry, i.e., the likelihood of future collaborations based on existing connections and collaborative history.}
\end{table*}\\ \vspace{-0.8cm} \vspace{-0.3cm}
\\
Temporal analysis revealed that older American actors clustered separately from contemporary Hollywood figures, with genre-specific clusters such as Hollywood TV and Hollywood Horror remaining isolated. However, as the interaction frequency reduced to 2\%, collaboration patterns expanded further, revealing significant partnerships. At 2\% interaction frequency shown by Fig. \ref{2percent}, the Hollywood cluster starts interacting with Old American, Italian, and Yugolovakian clusters, while the French cluster starts interacting with the Italian cluster. 

The dataset also highlighted co-acting relationships, with actors like Adoor Bhasi and Bahadur collaborating in over 187 films and Japanese actors Kijaku Otani and Matsunosuke Onoe co-acting in 146 films. The Louvain clustering technique resulted in a high modularity score of 90.8\%, indicating a strong, diverse network structure. The diversity of collaborations was further illustrated through high fit scores (94.9\%) and a notable variety of industry interactions. The analysis also identified Jackie Chan as the most versatile actor, adept at co-acting across different clusters, showcasing his unique role in bridging various industry segments. Link prediction estimated a high probability of future collaborations, particularly between actors like Gary Oldman and Marion Cotillard, with an 84\% likelihood of partnerships between Joseph Gordon-Levitt and Gary Oldman, as well as Gary Oldman and Marion Cotillard, as shown by Table \ref{linkprediction}.
\vspace{-0.5cm}
\begin{figure}[h]
    \centering
    \begin{minipage}{0.48\textwidth}
    \begin{tikzpicture}
        \begin{axis}[
            width=7.5cm,
            height=5.5cm,
            title={Trends of Movies vs. TV shows in OTT (2011-2020)},
            xlabel={Year},
            ylabel={Number of Releases},
            xmin=2011, xmax=2020,
            ymin=0, ymax=800,
            xtick={2012,2014,2016,2018,2020},
            ytick={0,100,200,300,400,500,600,700},
            legend pos=north west,
            ymajorgrids=true,
            grid style=dashed,
            legend style={font=\small}
        ]
        
        \addplot[
            color=blue,
            mark=*,
            ]
            coordinates {
            (2011,150)(2012,175)(2013,250)(2014,390)(2015,650)(2016,750)(2017,750)(2018,500)(2019,350)(2020,250)
            };
        
        \addplot[
            color=red,
            mark=*,
            ]
            coordinates {
            (2011,30)(2012,50)(2013,40)(2014,70)(2015,220)(2016,220)(2017,310)(2018,350)(2019,400)(2020,390)
            };
        
        \legend{\tiny Movies,\tiny TV shows}
        
        \end{axis}
    \end{tikzpicture}
    \caption{Comparative trends between TV shows and movies on OTT platforms from 2011-2020}
    \label{fig:ott-trends}
    \end{minipage}
    \hfill
    \begin{minipage}{0.48\textwidth}
    \begin{tikzpicture}
        \begin{axis}[
            width=7.5cm,
            height=5.5cm,
            title={Number of Movies per Cast Size},
            xlabel={Cast Size},
            ylabel={Frequency},
            xmin=0, xmax=50,
            ymin=0, ymax=1800,
            xtick={0,10,20,30,40,50},
            ytick={0,200,400,600,800,1000,1200,1400,1600,1800},
            ymajorgrids=true,
            grid style=dashed,
            bar width=1.5pt,
        ]
        
        \addplot[ybar, fill=blue!80!black] coordinates {
            (1,1700) (2,200) (3,220) (4,280) (5,375)
            (6,400) (7,500) (8,620) (9,620) (10,1180)
            (11,1440) (12,700) (13,490) (14,320) (15,200)
            (16,150) (17,120) (18,80) (19,60) (20,50)
            (21,40) (22,30) (23,25) (24,20) (25,15)
            (26,12) (27,10) (28,8) (29,7) (30,6)
            (31,5) (32,4) (33,3) (34,2) (35,2)
            (36,1) (37,1) (38,1) (39,1) (40,1)
            (41,0) (42,0) (43,0) (44,0) (45,0)
            (46,0) (47,0) (48,0) (49,0) (50,0)
        };
        
        \end{axis}
    \end{tikzpicture}
    \caption{Distribution of movies categorized by cast size, i.e., the frequency of movies featuring varying numbers of actors in the OTT industry.}
    \label{fig:movies-per-cast-size}
    \end{minipage}
\end{figure}\\ \vspace{-1cm}
\\
 Furthermore, our general analysis of the industry's growth trends indicated an increase in TV shows while a slight decrease of movies on OTT platforms from 2011 to 2020 (see Fig. \ref{fig:ott-trends}). Fig. \ref{fig:top-directors} and \ref{fig:top-actors} show that Rajiv Chilaka and Rahul Campos are the top directors who have directed the highest number of movies, while Anupam Kher and Rupa Bhimani are the actors who have acted in the highest number of movies. We also observed that traditional Bollywood actors appeared to be relatively isolated from the global cinema landscape, although exceptions existed. Bollywood actors such as Om Puri and Gulshan Grover were identified as the top Indian actors collaborating with Hollywood, reflecting the inter-industry connections forged. The analysis of movies and cast size as shown in Fig. \ref{fig:movies-per-cast-size} indicates that the majority of movies have a cast size of less than 10, with very few featuring a cast size above 20.
\vspace{-0.5cm}
\begin{figure}[h]
    \begin{minipage}[t]{0.48\textwidth}
        \centering
        \begin{tikzpicture}
            \begin{axis}[
                width=7.5cm,
                height=5.5cm,
                title={Top 5 Directors in OTT industry},
                xlabel={Directors},
                ylabel={Number of Movies},
                ymin=0, ymax=26,
                ytick={0,5,10,15,20,25},
                symbolic x coords={
                    {Rajiv Chilaka},
                    {Raul Campos},
                    {Jan Suter},
                    {Suhas Kadav},
                    {Marcus Raboy}
                },
                xtick=data,
                x tick label style={rotate=45,anchor=east},
                enlarge x limits=0.2,
                bar width=15pt,
                nodes near coords,
                nodes near coords align={vertical},
                ]
                
                \addplot[ybar,fill=blue!90!black] coordinates {
                    (Rajiv Chilaka,25)
                    (Raul Campos,20)
                    (Jan Suter,21)
                    (Suhas Kadav,18)
                    (Marcus Raboy,18)
                };
            \end{axis}
        \end{tikzpicture}
        \caption{The top 5 directors in the OTT industry by the number of movies directed}
        \label{fig:top-directors}
    \end{minipage}%
    \hfill
    \begin{minipage}[t]{0.49\textwidth}
        \centering
        \begin{tikzpicture}
            \begin{axis}[
                width=7.5cm,
                height=5.5cm,
                title={Movies Count of Top 5 Actors in OTT industry},
                xlabel={Actors},
                ylabel={Number of Movies},
                ymin=0, ymax=40,
                ytick={0,5,10,15,20,25,30,35,40},
                symbolic x coords={
                    {Anupam Kher},
                    {Rupa Bhimani},
                    {Tejwani},
                    {Om Puri},
                    {Takahiro Sakurai}
                },
                xtick=data,
                x tick label style={rotate=45,anchor=east},
                enlarge x limits=0.2,
                bar width=15pt,
                nodes near coords,
                nodes near coords align={vertical},
                ]
                
                \addplot[ybar,fill=blue!90!black] coordinates {
                    (Anupam Kher,38)
                    (Rupa Bhimani,32)
                    (Tejwani,27)
                    (Om Puri,35)
                    (Takahiro Sakurai,30)
                };
            \end{axis}
        \end{tikzpicture}
        \caption{Total number of movies produced by the top 5 actors in the OTT industry}
        \label{fig:top-actors}
    \end{minipage}
    \label{fig:ott-distribution}
\end{figure}\\ \vspace{-1cm}
\\
The study uncovered vibrant cross-language connections, with Chinese and South Korean actors leading a dynamic transnational cluster. Hollywood, ever adaptive, expanded its reach to embrace collaborations with Scandinavian, Spanish, Polish, Brazilian, and Nigerian films, showcasing its global flexibility. However, industries like Russian, Greek, Egyptian, and even Hollywood Porn remained more secluded, limited by fewer interactions.
\vspace{-0.5cm}
 \vspace{-0.4cm}

\begin{figure}[htbp]
  \centering
  \begin{minipage}[b]{0.48\textwidth}
    \centering
    \includegraphics[width=\textwidth]{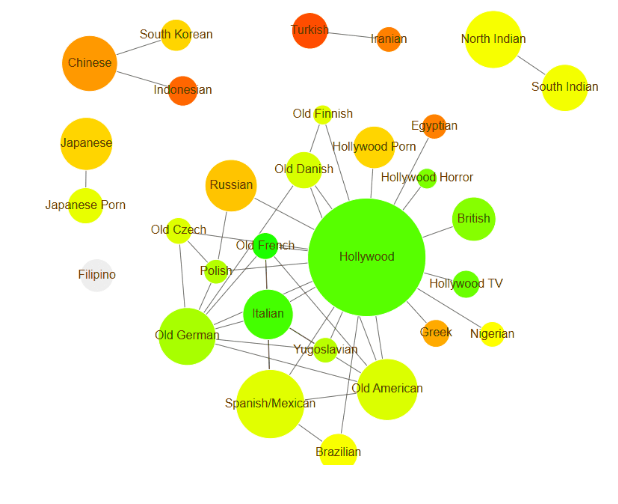}
    \caption{Visualization of cluster interconnection at 0.5\% frequency rate}
    \label{zeropointtwofice}
  \end{minipage}
  \hfill
  \begin{minipage}[b]{0.48\textwidth}
    \centering
    \includegraphics[width=\textwidth]{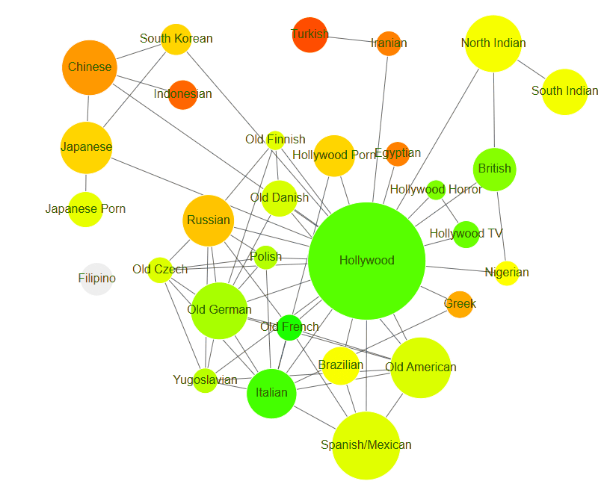}
    \caption{Visualization of cluster interconnection at 0.25\% frequency rate}
    \label{zeropointtwofive}
  \end{minipage}
\end{figure}

As shown by Fig. \ref{zeropointtwofice}, at the 0.5\% interaction frequency, our analysis unveiled that actors predominantly collaborated within their own actor groups, with the exception of a few collaborative clusters such as Turkish and Iranian actors, Indonesian actors interacting with the Chinese, and Hollywood's expansion to encompass Russian, Greek, and Egyptian actors. Finally, at the 0.25\% interaction frequency, shown by Fig. \ref{zeropointtwofive}, our findings demonstrated a more unified pattern, with almost every actor group engaging with another group to some extent. However, Filipino actors emerged as the most insular major actor group with limited intergroup interactions. This analysis provided valuable insights into the collaborative dynamics of Bollywood and its relationships with world cinema, offering a unique perspective on industry cohesion and isolation.
\vspace{-0.5cm}
\\ \vspace{-0.7cm}
\\
In summary, the network analysis, centrality measures, and cluster exploration allowed us to gain a comprehensive understanding of the film and OTT industry's intricate social networks. These findings offer valuable insights that can inform industry professionals, content creators, and streaming platforms, enabling them to make informed decisions about collaborations, partnerships, and content creation strategies in the ever-evolving entertainment landscape.

\section{Conclusion and Future Scope}\label{SCM}

In conclusion, our comprehensive analysis of social networks among movie actors and directors has provided valuable insights into the intricate web of collaborations in the modern film industry. By leveraging data from IMDb and Netflix, we identified key players, top actors, and directors who play central roles in the industry's network. We explored the increasing trend of TV shows on OTT platforms and unveiled patterns in centrality measures like degree, betweenness, and closeness centrality, highlighting influential actors who act as bridges between different clusters. Our cluster analysis revealed categorizations based on various criteria, emphasizing that actors predominantly collaborate within language groups, transcending national boundaries. Link prediction techniques offered valuable insights into potential future collaborations. The study also addressed key questions regarding the industry's isolation, indicating that it operates more on linguistic, temporal, and genre-based divisions than national boundaries. This research has significant implications for industry professionals, scholars, and enthusiasts, guiding decision-making processes in the dynamic world of film and entertainment.

Looking ahead, many exciting opportunities for exploration await. Temporal analysis could shed light on how collaboration patterns change over time. The influence of genre on actor collaborations and content analysis can provide deeper insights into storytelling and network dynamics. Machine learning models like SVM, deep learning models like ANN, and NLP techniques, also suggested by key word network analysis, can further increase knowledge and uncover insights of actor collaboration and also provide advanced link prediction and network analysis possibilities. The impact of globalization on the film industry's cross-country, cross-language collaborations remains largely unexplored. Investigating the economic implications and social media's role in shaping actor collaborations are promising research areas. These future directions will enrich our understanding of the film industry's social networks and evolving dynamics, offering valuable insights to entertainment stakeholders.

\end{document}